\documentstyle[12pt]{article}

\begin{document}
\titlepage
\title{
\begin{flushright}
Preprint PNPI-1947, January 1994
\end{flushright}
\bigskip\bigskip\bigskip
Critical  Composite Superconformal String}
\date{}
\author{V.A.Kudryavtsev\\ Petersburg Nuclear Physics
Institute\\ Gatchina, St.Petersburg 188350, Russia}
\maketitle

\vspace{1cm}

\begin{abstract}
In the paper the nilpotent conditions of BRST operator for
new superconformal string model were found. This string
includes anticommutation $2-d$ fields additional to the
standard Neveu-Schwarz superconformal set which carry quark
quantum numbers. In this case the superconformal symmetry
is realized by a non-linear way. In the superconformal
composite string new constraints for 1 and 1/2
conformal dimension should be added to the standard system
of Virasoro superalgebra constraints for 2 and 3/2
conformal dimensions. The number $N$ of the constraints and
numbers $D$ and $D'$ of bosonic and fermionic $2-d$ fields
are connected by a simple relationship $D'/2+D-3N-15=0.$
Also perspectives of the critical composite superconformal
string are discussed.
\end{abstract}

\newpage
Four dimensional string models \cite{1} derived from a
fermionization of compactification of $2-d$ bosonic fields
have essentially extended the field of critical string
under consideration. The next step forward in this
direction has been done in the recently proposed model of
 superfermionic superconformal string \cite{2}. In the
superfermionic string all the $2-d$ bosonic fields are
excluded and only fermionic ones are conserved. Such an
approach revealed some interesting possibilities for real
describing quark degrees of freedom  in the case of
hadrons, but it ceased to have a remarkable feature of
duality of amplitude  characteristic for all original
string theories \cite{3}. Synthesis of the classical
approach and the superfermionic model \cite{4} based upon
the generalization of multi-Regge vertices \cite{5} proved
to be more promising. The generalization allowed us to
build a realistic model of $\pi $-meson interactions
possessed spectrum of resonance states, Regge trajectories
and lengths of scattering in a good agreement with
experiment. In the composite superconformal model the most
attractive features of the original string approach
(superconformal symmetry and the duality) are conserved.

 In the present paper we find how one can obtain the
composite superconformal model with critical properties. As
known, the standard Neveu-Schwarz-Ramond string (NSR) is
critical in 10-d space-time. The composite string in
addition to the standard NSR fields \cite{6} $X_m\left(
\tau \right) $ and $ H_m\left( \tau \right) $ ,
$m=0,1,2\ldots ,d-1$ (a string coordinate and its
superpartner) has  a
supplementary set of 36 anticommutation fields of conformal
dimension $\frac 12$ \cite{2} in a superfermionic sector.
They are 8-component fields
$\Psi _{\alpha \beta }$ ( $\alpha $ --- a spinor index and
$ \beta $ --- isospinor one, $\alpha =1\ ,2,\ 3,\ 4$;
$\beta =1,2$) carring quantum numbers of an isospinor quark
with the spin $\frac 12$ and also 28 fields with quantum
numbers of currents built from these quarks. Among the 28
fields we have 10 isoscalar fields (4 vectors $\varphi _\mu
,$ 6 tensors $\xi _{\mu \nu }$) and 18 isovector fields (3
scalars $\eta _i$, 3 pseudoscalars $\upsilon _i$ and 12
pseudovectors $\chi _{\mu i}$ , $\mu =0,1,2,3;$ $i=1,2$).
These 36 components appear in the threelinear (in these
fields) scalar supergauge operator $G_r^f$ of Virasoro
algebra ($ G_r^f\equiv G_r^{\left( 0\right) }$ in
notations of ref.\cite{2}). The sum of $G_r^f$ and the
standard Neveu-Schwarz operator $G_r^{NS}=\oint \partial
\tau X_\mu H^\mu e^{-in\tau }$ gives a total supergauge
operator $G_r$ of the considered model
\begin{equation}
\label{1}G_r=G_r^f+G_r^{NS}
\end{equation}
which satisfies the anticommutation relations of the
standard Virasoro superalgebra $$ \begin{array}{l} \left\{
G_r,G_s\right\} =2L_{r+s}+\frac c3\left( r^2-\frac
14\right)\delta_{r,-s} \\ \left[ L_n,L_m\right] =\left(
n-m\right) L_{n+m}+\frac c{12}n\left( n^2-1\right) \delta
_{n,-m} \end{array} $$ \begin{equation} \label{2}\left[
L_n,G_r\right] =\left( \frac n2-r\right) G_{n+r}
\end{equation}

where
\begin{equation}
\label{3}c=\frac{3d}2+18\qquad .
\end{equation}

Wave vectors of physical states $\mid \Phi \rangle $ satisfy the
corresponding gauge conditions
\begin{equation}
\label{4}
\begin{array}{l}
G_r\mid \Phi \rangle =0\qquad r>0 \\
L_n\mid \Phi \rangle =0\qquad n>0 \\
\left( L_0-a\right) \mid \Phi \rangle =0
\end{array}
\end{equation}
where a constant $a=\frac 12$ in the NSR model.

For the standard NS superstring the critical value of $c,$
which provides unitarity in loop corrections, is equal to
15 (that is $d=10$ in the absence of the superfermionic
sector). For $d\geq 0$ in the composite superconformal
model with the conditions $\left( \ref{4}\right) $ the
value $ \left( \ref{3}\right) $ is not critical and,
consequently, the unitarity on loop level is lost. However,
as we shall see, one can achieve the critical supergauge
algebra of the constraints by having involved additional
constraints
\begin{equation}
\label{5}
\begin{array}{l}
\nu _r^{\left( l\right) }\mid \Phi \rangle =0\qquad r>0 \\
J_n^{\left( l\right) }\mid \Phi \rangle =0\qquad n>0
\end{array}
\begin{array}{c}
l=1,2,\ldots N \\
\end{array}
\end{equation}
where $\nu _r^{\left( l\right) }$ some $N$ linear
combinations of fields from the considered above 36
components of the superfermionic sector and $ J_n^{\left(
l\right) }=\left\{ G_{n-r},\nu _r^{\left( l\right)
}\right\} $ are their associated currents. Certainly, it
also means the requirement of invariance of string vertices
under the new constraints $\left( \ref{5} \right) .$ The
incorporation of the new gauge constraints $\left( \ref{5}
\right) $ produces new items which depend on new ghost and antighost
operators $d_n^{\left( l\right) },\ell _n^{\left( l\right) },\delta
 _r^{\left( l\right) }$ and $\alpha _r^{\left( l\right) }$
in BRST charge $Q$  according to commutation relations
for $\nu _r^{\left( l\right) }$ and $ J_n^{\left( l\right)
}$
\begin{equation} \label{6} \begin{array}{l} \left[
J_n^{\left( i\right) },J_m^{\left( j\right) }\right]
=f_{ijk}J_{n+m}^{\left( k\right) }+g_{ij}n\delta _{n,-m} \\
\\
 \left[ L_n,J_m^{\left( i\right) }\right]
=-mJ_{n+m}^{\left( i\right) } \\ \\ \left\{ G_r,\nu
_s^{\left( i\right) }\right\} =J_{r+s}^{\left( i\right) }
\\ \\ \left[ L_n,\nu _s^{\left( i\right) }\right] =\left(
-\frac n2-s\right) \nu _{n+s}^{\left( i\right) } \\ \\
\left[ J_n^{\left( i\right) },\nu _s^{\left( j\right)
}\right] =t_{ijl}\nu _{n+s}^{\left( l\right) } \\ \\
\left\{ \nu _s^{\left( i\right) },\nu _r^{\left( j\right) }\right\}
=h_{ij}\delta _{r,-s}
\end{array}
\end{equation}

If the corresponding operator of the BRST charge $Q$ for
the composite superconformal string is written in full,
then we have
\begin{eqnarray} \label{7}
Q&=&\sum_n L_{-n}c_n+\sum_r G_{-r}\gamma_r
+\sum_n J_{-n}^{(i)} d_n^i+\sum_r\nu
_{-r}^{(j)}\delta_r^{(j)} \nonumber\\
&-& \frac 12 \sum_{n,m}( m-n):c_{-m}c_{-n}b_{m+n}:
-\sum_{n,r}\left( \frac{3n}2+r\right) :c_{-n}\beta_{-r}
\gamma _{r+n}: \nonumber\\
&-&\sum_{r,s}\gamma _{-r}\gamma_{-s}b_{r+s} -\frac 12
\sum_{i,j,k,n,m}f_{ijk}:d_{-n}^{(i)}d_{-m}^{(j)}
\ell_{n+m}^{(k)}: \nonumber \\
&+&\sum_{n,m}(n+m) c_{-n}\ell _{-m}^{(i)}d_{m+n}^{(i)}
-\sum_{r,m}\gamma_{-r}\ell_{-m}^{(i)}\delta_{r+m}^{(i)}
\nonumber \\ &-&
\sum_{i,r,s}(r+s)\gamma_{-r}\alpha_{-s}^{(i)}d_{r+s}^{(i)}
-\sum_{n,s}\left( \frac n2+s\right) c_{-n}\alpha_{-s}^{(i)}
\delta_{n+s}^{(i)}\nonumber \\
&-&\sum_{i,l,j}t_{ijl}d_{-n}^{(i)}\alpha_{-s}^{(j)}
\delta_{n+s}^{(l)} +\sum_i f^{(i)}d_0^{(i)}-ac_0
\end{eqnarray}
where $$ \left\{ c_n,b_m\right\}
=\delta _{n,-m}\qquad \left[ \beta _r,\gamma _s\right]
=\delta _{r,-s} $$ $$ c_n^{+}=c_n\qquad
b_m^{+}=b_{-m}\qquad \ell _n^{\left( i\right) +}=\ell
_{-n}^{\left( i\right) }\qquad d_n^{\left( i\right) +}=d_{-n}^{\left(
i\right) }
$$
$$
\beta _r^{+}=-\beta _{-r}\qquad \gamma _r^{+}=\gamma
_{-r}\qquad \alpha _r^{\left( i\right) +}=-\alpha
_{-r}^{\left( i\right) }\qquad \delta _s^{\left( j\right)
+}=\delta _{-s}^{\left( j\right) } $$ $$ \left[ \alpha
_r^{\left( i\right) },\delta _s^{\left( j\right) }\right]
=\delta _{i,j}\delta _{r,-s}\qquad \left\{ \ell _n^{\left(
i\right) },d_m^{\left( j\right) }\right\} =\delta
_{i,j}\delta _{n,-m} $$

Now, as for BRST approach in the original string theories
one should find conditions that the BRST charge $Q$ is
nilpotent, which means that the superalgebra of the
gauge constraints is critical one. As a result of this
 nilpotency $Q^2=0$ anomal terms are absent in the
commutation relations $ \left( \ref{8}\right) $ and
$\left( \ref{9}\right) $.  \begin{equation} \label{8}\left[
L_n^{quant},L_m^{quant}\right] =\left( n-m\right)
L_{n+m}^{quant}
\end{equation}

\begin{equation}
\label{9}\left\{ G_r^{quant},G_s^{quant}\right\} =2L_{r+s}^{quant}
\end{equation}

 where $L_n^{quant}=\left\{ b_n,Q\right\} $ and
$G_s^{quant}=\left[ \beta _s,Q\right] $

 From disappearance of anomalies in $\left( \ref{8}\right)
$ and $\left( \ref {9}\right) $ we obtain two relationships
\begin{equation}
\label{10}D+\frac{D^{\prime }}2-3N-15=0
\end{equation}
\begin{equation}
\label{11}a=\frac 12
\end{equation}

where $D$ is a number of bosonic fields and $D^{\prime }$
is a number of fermionic fields. In this case we have
$D=d,$ and $D^{\prime }=d+36$.

 The relationship $\left( \ref{10}\right) $ becomes readily
available when we know contributions of ghosts of conformal
dimension $\jmath $ in the anomaly $\left( \ref{8}\right)
$: $\left( \pm 1\right) \left( 3\left( 2\jmath -1\right)
^2\right.$ \\ $\left.-1\right) $. The sign $\pm $ is
defined by statistics of the ghost. We have $\left(
-1\right) 26$ for $\jmath =2$, $\left( +1\right) \left(
+11\right) $ for $\jmath =\frac 32$, $\left( -1\right)
\left( +2\right) $ for $\jmath =1$, and $\left( +1\right)
\left( -1\right) $ for $ \jmath =\frac 12$ that allows us
to obtain $$ -26+11-2N-N=-15-3N\ .  $$

A concrete method of construction of string vertices and
amplitude of the composite superconformal string \cite{4}
implies that both "gluonic" Neveu-Schwarz sector and
"quark" (superfermionic) sector are individually
critical. So we choose here $d=10$ and $N=6$. Furthermore,
it appears that the conditions of absence of anomalies
don't insure $Q^2=0$ because of anomalies in the
commutation relations $\left\{ \nu _s^{\left( i\right)
},\nu _r^{\left( j\right) }\right\} $ and $\left[
J_n^{\left( i\right) },J_m^{\left( j\right) }\right] $ in
$\left( \ref{6}\right) $. One can check that the nilpotent
condition $Q^2=0$ requires additional constraints
\begin{equation}
\label{12}
\begin{array}{c}
f^{\left( i\right) }=0\qquad h_{ij}=0\qquad \sum_jf_{ijj}=0 \\
\\
\sum_jt_{jij}=0\qquad \sum_jt_{ijj}=0
\end{array}
\end{equation}

The disappearance of coefficients $h_{ij}$ is possible due
to a signature of Minkowski metric and existence of the
components $\varphi _0,\xi _{0\mu },\chi _{0i}$ in the
spectrum of the superfermionic sector \begin{equation}
\label{13}
\begin{array}{c}
\left\{ \left( \varphi _\mu \right) _r,\left( \varphi _\nu \right)
_s\right\} =-g_{\mu \nu }\delta _{r,-s} \\
g_{\mu \nu }=diag\left( 1,-1,-1,-1\right)
\end{array}
\end{equation}

Choose, as $\nu ^{\left( l\right) }\quad l=1,2,\ldots N$ ,
six light-like components $\varphi _0+\varphi _3,\ \xi
_{01}+\xi _{31},\ \xi _{02}+\xi _{32},\ \chi _{0i}+\chi
_{3i}\quad \left( i=1,2,3\right) $. By choosing so one can
obtain $Q^2=0$ that is the desired condition. It is
remarkable that in this case, as for the original critical
string ( the bosonic string $ \left( d=26\right) $ and the
superstring $\left( d=10\right) $ ), an elimination of
spurious states with null norm has occured together with
their conjugate ones. This leads to vanishing both null and
longitudinal components. In other words the elimination of
$\varphi _0+\varphi _3,\ \xi _{01}+\xi _{31},\ \xi
_{02}+\xi _{32},\ \chi _{0i}+\chi _{3i}$ occurs together
with their conjugate $\varphi _0-\varphi _3,\ \xi _{01}-\xi
_{31},\ \xi _{02}-\xi _{32},\ \chi _{0i}-\chi _{3i}$.
Hence, $\varphi _0,\ \varphi _3,\ \xi _{01},\ \xi _{31},\
\xi _{02},\ \xi _{32},\ \chi _{0i},\ \chi _{3i}$ fall out
from spectrum of physical states. Moreover, in the critical
composite superconformal model the elimination of spurious
states appears to be still more effective by virtue of the
nontrivial commutation relations $\left( \ref{6}\right) $
for these components and currents connected with them. It
is interesting that in the string amplitude \cite{4} one
can choose as light--like $\left( k_i^2=0\right) $ all
momenta along the quark lines of dual quark
diagram\cite{7}. Since the superfermionic sector in the
amplitudes \cite{4} is chosen for each quark lines,
this is in its turn achieving a necessary light direction
for the considered above conditions $\varphi _0+\varphi
_3=k_\mu \varphi ^\mu $ and so on, and hence we insure
Lorentz invariance of this approach.

The analysis which has been out here corresponds to the
first generation ($u$ and $d$ quarks of the Standard
Model).  Other quantum numbers which correspond to $s,c,b$
and $t$ quarks and baryon number will be a subject of a
following paper. It is to be noted that the number of the
fermionic fields exceeds the number of bosonic ones in the
spectrum of the composite superconformal string, and this
fact together with absence of tachions and dilatons
 inevitably has  to lead to finiteness of oriented (only
possible here) one-loop (and, perhaps, multi-loop)
diagrams. As we know the leading divergence of integral in
the Chan variable $x$ for the one-loop diagram is defined
by the expression $\exp \left( \frac{\pi ^2}6\left(
\widetilde{D}-\widetilde{D}^{\prime }\right) \right) \frac
1{1-x}$ in the limit $x\rightarrow 1$ where
$\widetilde{D},\widetilde{D}^{\prime }$ -full numbers of
boson and fermion fields correspondingly including ghost
fields. In the case of $\widetilde{D}\ <\
\widetilde{D}^{\prime }$ and $ x\leq 1$ this integral
converges.\\

The author would like to thank E.N.Antonov,
A.P.Bukhvostov, G.V.Fro- lov, L.N.Lipatov for useful
discussions.\\

Work supported in part by Russian Fund of
Fundamental Research under Contract No.93-02-3864

\end{document}